\documentclass[
reprint,
nofootinbib,
superscriptaddress,
aps, 
aas,
longbibliography,
color,
tabulary,
floatfix,
twocolumn
]{revtex4}

\usepackage{latexsym,epsfig,amssymb, amsmath, nicefrac, slashed, color, graphicx,amsfonts}
\usepackage{ulem}
\usepackage{hyperref}
\pdfoutput=1

\usepackage{float}

\renewcommand{\bf}[1]{\textbf{#1}}

\begin{document}

\title{The Special Galileon as Goldstone of Diffeomorphisms}

\author{Diederik Roest}

\affiliation{Van Swinderen Institute for Particle Physics and Gravity, University of Groningen, Nijenborgh 4, 9747 AG Groningen, The Netherlands}

\begin{abstract} \noindent
The special Galileon stands out amongst scalar field theories due to its soft limits, non-linear symmetries and scattering amplitudes. This prompts the question what the origin of its underlying symmetry is. We show that it is intimately connected to general relativity: the special Galileon is the Goldstone mode of the affine group, consisting of linear coordinate transformations, analogous to the dilaton for conformal symmetries. We construct the corresponding metric, and discuss various relations to gravity, Yang-Mills and the non-linear sigma-model.
\end{abstract}
	
\maketitle

\section{Introduction}

Symmetries in various guises are central to many fundamental theories of Nature. Gravity is governed by diffeomorphisms, while Yang-Mills (YM) interactions are dictated by gauge symmetry; both are the essentially unique Lorentz-invariant  spin-2 and spin-1 interactions. Scalar field theories, while less constrained by Lorentz symmetry, can also realise different symmetries, with applications ranging from QCD's pions to inflation's shift symmetry. Amongst these, what is the highest degree of symmetry that can be attained for a single scalar? A natural answer to this question is the special Galileon.

A first argument for this answer stems from a systematic analysis of the soft limits of scalar theories \cite{softlimits1,softlimits2}. For a shift symmetric scalar field, the amplitude of any process vanishes (the `Adler zero') in the soft limit of zero momentum for an external state. At the lowest order in derivatives, i.e.~with $(\partial \phi)^n$ interactions, there is a unique way to make this limit as soft as $A \sim p^2$ by tuning to the specific Dirac-Born-Infeld (DBI) interactions. With exchange and contact diagrams cancelling each other's $A \sim p$ contributions, this is the first example of an `exceptional' scalar field theory\footnote{For multiple scalars one can think of the non-linear sigma model (NLSM) as an exceptional field theory at the two-derivative level, with the Adler zero requiring cancellations between   diagrams.}. At the next order in derivatives, with $(\partial \phi)^2 (\partial \partial \phi)^{n-2}$ interactions, one finds the Galileon interactions with the same soft limit as DBI. Again, there is an exceptional theory with tuned interactions and soft limit $A \sim p^3$; this is the special Galileon.

Underlying these soft limits are non-linearly realised symmetries. The Adler zero follows from the shift symmetry $\delta \phi = a$. Both DBI and the Galileon interactions inherit their soft limit from the linear shift $\delta \phi = b_\mu x^\mu + \ldots$, suppressing possible field dependent terms. Finally, the special Galileon allows for an additional transformation with a symmetric and traceless generator of the form $\delta \phi = s_{\mu \nu} x^\mu x^\nu + \ldots$ \cite{Hidden}. It has been shown, both from the amplitude side and by an algebraic classification \cite{Werkman}, that there are no further possibilities: the SG is the softest and most symmetric scalar theory.

The special Galileon also makes an interesting appearance in the double copy relations of scattering amplitudes. Initially proposed for gravity and Yang-Mills \cite{Bern}, there is now a web of relations between seemingly different theories, indicated in figure 1. For all theories indicated, the amplitudes can be written as a sum over trivalent exchange diagrams $i$ whose contribution can be factorised into $f_i f_i / D_i$, with colour or kinematical factors $f \in (c,n,r)$ satisfying Jacobi-like identities and propagator structures $D_i$ (see \cite{Huang, Cheung} for reviews). 

Colour-kinematic duality relates these theories by simply replacing these factors. For instance, at the top of the triangle one has a biadjoint scalar theory (BS) with $\phi^{a \bar a}$ in the adjoint of $SU(N) \times SU(M)$ and with cubic interactions, whose amplitudes factorise into a sum over two colour factors $c_i$ that feature the structure constants. Moving down, the amplitudes take the same factorised from, with colour factors replaced by kinematical ones. Yang-Mills and general relativity (GR) follow from successively replacing colour with kinematics $n$. Alternatively, one finds non-linear sigma models and the special Galileon (SG) when using the alternative kinematic factor $r$ instead. 

\begin{figure}[t!]
\center
	\includegraphics[width=0.23\textwidth]{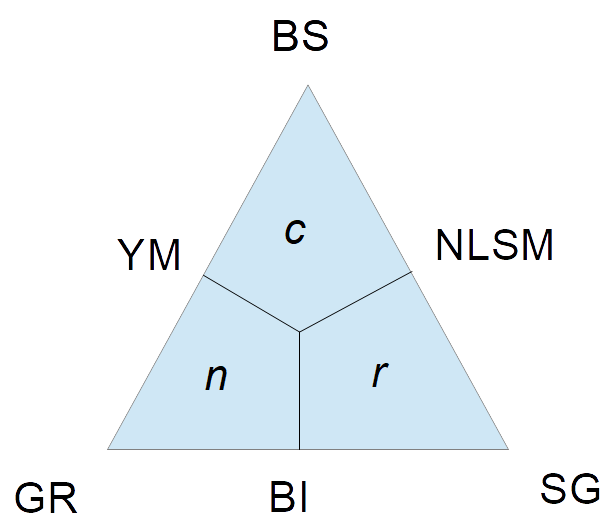}
	\caption{\it A web of theories whose amplitudes are the products of colour ($c$) or kinematical ($n,r$) factors subject to Jacobi-like identities.}
	\label{fig:cartoon} 
	 \vspace{-0.5cm}
\end{figure}

Various relations have been pointed out between the amplitudes \cite{CHY, unifying} and dimensional reductions \cite{pions} of these theories. As we will recap below, the NLSM arise as the non-linear realisation of the global part of YM gauge symmetries. This prompts the question whether there is a similar relation between GR and the special Galileon: does the latter arise as the non-linear realisation of a restricted subset of diffeomorphisms? This would tie in nicely with the similarities between gravity and this specific field theory, with both adhering to the equivalence principle and soft limit theorems \cite{Equivalence}. 

In this paper we will put this suspicion on firm footing: it will be shown that the special Galileon arises as the Goldstone boson of special linear transformations. This identification allows for a novel formulation of the theory in terms of a simple covariant metric, with interesting links to the metrics defining other theories in figure 1.

\section{Goldstones}

We first recap the relation between Yang-Mills and the non-linear sigma model, halfway down figure 1. Yang-Mills vectors transform under the gauge symmetry as
 \begin{align}
  A_\mu \rightarrow U A_\mu U^{-1} + U \partial_\mu U^{-1} \,, \label{YM-transf}
 \end{align}
thus determining the gluon self-interactions as well as those with particles that carry a colour charge. 

A natural question is whether there is a `smaller' theory inside this structure - invariant under only a subset of the full gauge symmetry and thus doing away with the necessity of vector bosons. Indeed there is: one can consistently restrict the gauge symmetry to global transformations (picking out the physical part of the furthermore redundant gauge symmetry). Furthermore, this global symmetry is non-linear realised by the configuration
 \begin{align}
  A_\mu = U \partial_\mu U^{-1} \,, \quad U = \exp (i \phi^a (x) T_a / \Lambda) \,, \label{gauge-configuration}
 \end{align}
in terms of a Lie algebra element $U$ with scalar fields $\phi^a$ and symmetry breaking scale $\Lambda$. This configuration is covariant resp.~invariant under two independent global transformations, $U \rightarrow L U R$. The scalar fields transform non-linearly under the diagonal combination $L = R$ (and linearly under the anti-diagonal combination $L = R^{-1}$); hence they are the Goldstone bosons of the spontanously symmetry breaking $G \times G \rightarrow G$.

\medskip

Turning to general relativity, the relevant gauge symmetry is generated by diffeomorphisms $x^\mu \rightarrow x^\mu - \xi^\mu$. Remarkably, this infinite-dimensional symmetry group follows from the closure of two-finite-dimensional groups: the conformal group and the affine group consisting of linear transformations \cite{Ogievetsky}. Both groups are finite-dimensional extensions of the Poincare group and natural restrictions of the full diffeomorphisms  (any further extension would generate the entire set of diffeomorphisms), with corresponding Goldstone modes. Indeed, linear transformations on the coordinates, $x^\mu \rightarrow \Lambda^\mu{}_\nu x^\nu$, are the direct counterpart of global YM transformations. Similar to the YM case, where the vector bosons transform  \eqref{YM-transf} covariantly under the global symmetry, the Christoffel symbols $\Gamma_{\mu \nu}{}^\rho$ transform covariantly under diffeomorphisms restricted to linear transformations. This strongly suggests to think of the conformal and the affine groups as the GR counterparts of global symmetries in YM.

Can one also construct non-linear realisations in terms of Goldstone bosons? Starting with the conformal group with dilations that act as\footnote{Parameters and generators will be indicated with lower- and uppercase letters, in this case $d$ for dilations $D$.} $\xi^\mu = - d / L \cdot x^\mu$,  the Minkowski metric transforms with a specific dilation weight, $\delta \eta_{\mu \nu} = - d / L \cdot \eta_{\mu\nu}$. Dilations are therefore a symmetry of the equivalence class of conformally flat metrics 
 \begin{align}
    g_{\mu \nu} = e^{- 2 \pi / L} \eta_{\mu \nu} \,. \label{metric-aff}
  \end{align}
The dilation weight of the metric can be absorbed into the transformation of the dilaton\footnote{Transformations are given in the passive form, with the active form being $\delta \phi + \xi^\mu \partial_\mu \phi$ and invariant coordinates.}, $\delta \pi = d$. This shows that the conformally flat metric transforms covariantly under the non-linear transformation, i.e.~with $\delta g_{\mu\nu} = \mathcal{L}_{\xi} g_{\mu\nu}$; any diffeomorphism-invariant object constructed out of this metric will therefore have this non-linear symmetry.

As is well-known, this metric has an additional non-linear symmetry of special conformal transformations,
  \begin{align}
   \xi^\mu =  \frac{(x^\mu k^\nu - \tfrac12 k^\mu  x^\nu) x_\nu}{L} \,,
  \end{align}
provided the dilaton transforms with a linear shift, $\delta \pi = k_\mu x^\mu$. Remarkably, the non-linear realisation of special conformal transformation does not require a separate Goldstone, and instead can be realised on the dilaton only. The would-be Goldstone mode for the special conformal symmetry is eliminated by imposing inverse Higgs constraints \cite{Ivanov, Low}, highlighting a crucial distinction between internal and space-time symmetry groups. 

\medskip

The second possibility is the affine group with coordinate transmations of the form $s^\mu{}_\nu x^\nu$ in addition to translations and Lorentz transformations. The parameter $s$ is symmetric when written with indices lowered with the Minkowski metric, and hence complementary to the anti-symmetric Lorentz parameters. While the affine parameters are not isometries of the Minkowski metric, the combination
 \begin{align}
  g_{\mu \nu} = \eta_{\mu \nu} + \frac{\partial_\mu \partial_\nu \phi}{\Lambda^3} \,, \label{metric-aff}
 \end{align}
does transform covariantly under
linear coordinate transformations provided they are accompanied by a scalar transformation,
 \begin{align}
  \delta \phi = s_{\mu\nu} x^\mu x^\nu \,, \quad
  \xi^\mu = \frac{s^\mu{}_\nu x^\nu}{\Lambda^3} \,. \label{SG-transf}
\end{align}
The symmetry breaking scale $\Lambda$ will be kept implicit in what follows.

The above metric in terms of $\phi$ therefore has an enhanced symmetry: it transforms covariantly under \eqref{SG-transf}.  This is similar to the non-linear transformation $U \rightarrow L U L$ in the YM example. The scalar field $\phi$ can therefore be seen as the Goldstone field of linear coordinate transformations. In the next sections, its truncation to special linear transformation and the relation to the special Galileon will be discussed.
 
One might be surprised by the mismatch between the tensor nature of the broken symmetries and the scalar Goldstone. Indeed, one could also have introduced a tensor Goldstone mode via $g_{\mu\nu} = \eta_{\mu\nu} + h_{\mu\nu}$ for the affine group, see e.g.~\cite{Isham, Borisov}. Similar to the YM gauge field \eqref{YM-transf}, it has a non-linear transformation whose passive form is
 \begin{align}
 \delta h_{\mu \nu} =  2 \partial_{(\mu} \xi_{\nu)} \,,
 \end{align}
under diffeomorphisms $\xi^\mu$. Remarkably, for linear coordinate transformations this transformation law is compatible with imposing the condition $h_{\mu \nu} = \partial_{(\mu} A_{\nu)}$ (that should be seen as the GR analogue of \eqref{gauge-configuration}), with transformation
 \begin{align}
 \delta A_\mu = 2 s_{\mu\nu}  x^\nu \,, \quad \xi^\mu = {s^\mu{}_\nu x^\nu} \,.
 \end{align}
In turn, the vector can be truncated to its longitudinal mode $A_\mu = \partial_\mu \phi$, leading to the scalar transformation \eqref{SG-transf} under linear diffeomorphism. These matryoshka-like relations are similar to the inverse Higgs condition in the conformal case; however, in the affine case they introduce a new vector and scalar field that are not directly associated to a spontaneously broken diffeomorphism. Instead, due to the double derivative, the metric with affine symmetry is also manifestly invariant under the shift and Galileon symmetry. 

The above demonstrates that, while the NLSM naturally follows from as a non-linear realisation of restricted gauge transformations, for GR there are actually two possibilities with non-linearly realised conformal and affine symmetries on the conformally flat metric  and on \eqref{metric-aff}. An interesting difference is that the gauge configuration \eqref{gauge-configuration} leads to a vanishing field strength, while both metrics have non-vanishing curvatures. We will return to this point when discussing the longitudinal modes of massive Lorentz representations.
 
\section{Dynamics}

Given the covariance of the metrics under the conformal or affine symmetries, the field equations for these scalar fields have to involve geometric objects such as the measure and the Riemann curvature. For the dilaton, the first relevant term in a derivative expansion is the Ricci scalar, given by
 \begin{align}
  R = - 6  e^{3 \pi / L} \Box e^{- \pi / L}  \,,
\end{align}
in four dimensions. Via a simple field redefinition, this is just proportional to $\Box \phi$ and the dilaton theory is therefore a free theory at lowest order in derivatives. Higher-derivative contributions can be added but these are optional; no self-interactions of the dilaton are required by the non-linear conformal symmetry. Instead, adding a cosmological constant amounts to including a quartic $\lambda \phi^4$ interaction for this free theory. This option was singled out as the unique non-derivative interaction that is recursively constructible in the soft bootstrap programme \cite{Elvang}. 

Aside self-interactions, the non-linear symmetry forces all matter to be minimally coupled with the conformal metric and hence leads to the equivalence principle. Interestingly, this theory was proposed as a precursor to GR in 1913 \cite{Nordstrom}, with the correct $1/r^2$ gravitational force in the Newtonian limit. Its relativistic corrections, however, induce a lag of the perihelion of planetary orbits, in contrast to GR's advance and Mercury's observed orbit. See also \cite{Sundrum} for a modern discussion.

\medskip

The situation is different for the special Galileon. In this case, the cosmological constant gives rise to the kinetic term plus self-interactions, with curvature terms providing higher-derivative corrections. At lowest order in derivatives, the field equations are therefore uniquely given by $\det(g)=1$, the condition of unimodular gravity. This leads to the field equation for the special Galileon\footnote{Alternatively, by taking a function of this condition one can manipulate it into $\rm{tr} ( \ln( 1 + \Pi))=0$ involving only single traces.},
 \begin{align}
  \sum_{i =1}^D T_i  = 0 \,, \quad   T_1 = [ \Pi ]  \,, \quad T_2 = \tfrac12 ( [ \Pi^2 ] - [ \Pi ]^2) \,,
 \end{align}
in terms of the total derivatives $T_i$ that arise at every order in the expansion of $\det(1+\Pi)$, where $\Pi \equiv \partial \partial \phi$ and $[]$ denote traces. 

At every order in fields, these terms are the Galileon interactions that are invariant under $\phi \rightarrow \phi + a + b_\mu x^\mu$ and have second order field equations \cite{Nicolis}; they are the Wess-Zumino terms of this symmetry \cite{Goon}. Moreover, the overall coefficient of every term is exactly the necessary one to enhance to the special Galileon symmetry \cite{Hidden}. This theory therefore needs specific self-interactions to realise its symmetry, similar to GR. 

The dynamics of the special Galileon imposes a condition on the special Galileon transformation \eqref{SG-transf}: its parameters $s_{\mu\nu}$ must be traceless\footnote{{Note that this does not imply that our original Goldstone mode, $h_{\mu \nu} = \partial_\mu \partial_\nu \phi$, has a vanishing trace; instead its trace is given in terms of the traceless part.}}. Indeed, one can see from the field equations starting with $\Box \phi + ...$ that these are only invariant under shifts with traceless quadratic terms. Equivalently, the unimodularity condition is invariant under the diffeomorphisms $\chi^\mu = s^{\mu \nu} x_\nu$ {\it provided} one restricts to symmetric and traceless parameters, i.e.~to special linear transformations; indeed the unimodularity condition is equivalent to imposing $\nabla_\mu \xi^\mu = 0$ for these transformations. 
  
\section{Contact}

We now turn to different representations of these theories in order to make contact with related results in the literature. These can arise due to field redefinitions, which in general map the fields (and the coordinates) onto themselves. Such relations form the natural generalisation of coordinate transformations and are known as point transformations.  Specifically for a single scalar field, however, one can perform a more general  field redefinition, mapping the set of coordinates, scalar and its derivative onto itself. These are known as contact (or local Lie-B{\"a}cklund) transformations. 

An important example of a contact transformation relates the conformal and Anti-de Sitter representations of the $SO(3,2)$ algebra \cite{Bellucci, Creminelli}, and brings one to the metric
 \begin{align}
  g_{\mu\nu} = e^{- 2 \pi/L} \eta_{\mu\nu} + \alpha \partial_\mu \pi \partial_\nu \pi \,.
 \end{align}
In terms of the new coordinates and field, the non-linear dilation acts in the same way, while the special conformal transformation now corresponds to a diffeomorphism
 \begin{align}
   \chi^\mu = \frac{(- x^\mu k^\nu + \tfrac12 k^\mu  x^\nu) x_\nu}{L} + \frac{\alpha L}{2} ( e^{2 \pi/L} - 1) k^\mu \,, \label{spec-conf-2}
 \end{align}
while at the same time the dilaton transforms as $\delta \pi = k_\mu x^\mu$. 

The special conformal generators now close as $[ K_\mu, K_\nu] = 8 \alpha M_{\mu\nu}$. All values of $\alpha$ are related via contact transformations, and hence equivalent. However, a crucial difference between the conformal and AdS representations is the appearance of $\pi$ in the diffeomorphism \eqref{spec-conf-2}. This highlights the fact that only the conformal representation has a four-dimensional interpretation as spontaneously broken diffeomorphisms. Instead, for $\alpha \neq 0$ this necessarily involves a brane with location $\pi$ in an extra dimension, as outlined in \cite{Reunited}; from the 4D perspective, this is a point (rather than a coordinate) transformation.

As a side note, taking the $L \rightarrow \infty$ limit of the AdS representation yields the Dirac-Born-Infeld theory with metric 
 \begin{align}
  g_{\mu \nu}  = \eta_{\mu \nu} + \alpha \partial_\mu \pi \partial_\nu \pi \,,
 \end{align}
which non-linearly realises 5D Poincar\'{e}. In this limit, the transformation \eqref{spec-conf-2} simplifies to 
 \begin{align}
  \delta \pi =  k_\mu x^\mu \,, \quad \chi^\mu = \alpha \pi k^\mu \,.
 \end{align}
While this follows from a geometric 5D construction, no such interpretation is possible in four dimensions. Moreover, the sign of $\alpha$ is physically relevant after this limit, with the positive case leading to $ISO(4,1)$ and negative to $ISO(3,2)$; of these, only the former satisfies the positivity bounds of \cite{Dubovsky} and thus allows for a sensible UV completion. The failure of the latter (`anti-DBI') to satisfy these UV constraints can be directly related to the two times of its geometric origin. Finally, for DBI there is a straightforward generalisation to include multiple scalars, corresponding to a four-dimensional brane with higher co-dimension.

\medskip

Turning to the special Galileon, the contact transformations known as Galileon duality \cite{GD} provide a mapping between different Galileon invariants $T_i$. They map the metric with non-linear affine symmetry \eqref{metric-aff} onto
 \begin{align}
  g_{\mu \nu} = \eta_{\mu \nu} + \tilde \alpha \partial_\mu \partial_\nu \phi + \tilde \beta  \partial_\mu \partial_\rho \phi \partial_\nu \partial^\rho \phi \,. \label{metric-gen}
 \end{align}
with parameters given by $\tilde \alpha = 1 + 2 \epsilon$ and $\tilde \beta = \epsilon + \epsilon^2$. Under Galileon duality, the symmetry transformations become
 \begin{align}
   \delta \phi & = s_{\mu \nu} x^\mu x^\nu - \tilde \beta s_{\mu \nu} \partial^\mu \phi \partial^\nu \phi \,, \notag \\
    \xi^\mu & = \tilde \alpha  s^{\mu \nu} x_\nu + 2 \tilde \beta s^{\mu \nu} \partial_\nu \phi \,,
 \end{align}
Note that these diffeomorphisms no longer preserve the unimodularity condition; instead, the dynamical condition becomes\footnote{One can think of this as a unimodularity condition $\det( \tilde \eta^{\mu \nu} g_{\nu \rho})=1$ for $g_{\mu \nu}$ with respect to a non-canonical flat background metric $\tilde \eta_{\mu\nu} = (1 + \epsilon \Pi)^2$ (similar to \eqref{MG-metric}).}
 \begin{align}
 \det(g) = \det(1 + \epsilon \Pi)^2 \,, \label{unimod-gen}
 \end{align}
which translates into 
 \begin{align}
   \sum_{i=1}^{D} ((1 + \epsilon)^i - \epsilon^i) T_i = 0 \,,
 \end{align}
as field equation for the special Galileon.

These transformations were first identied for general values of the parameters $\tilde \alpha, \tilde \beta$ in \cite{Hidden} (in active form, without covariant metric and diffeomorphisms) as the special Galileon symmetry that explains the $A \sim p^3$ soft limit of this theory found in \cite{Cheung}. 

How is this more general form related to the Goldstone of affine transformations of the previous sections? Galilean duality shifts the two parameters with $2 \epsilon$ and $\epsilon \tilde \alpha + \epsilon^2$, respectively \cite{Hidden}. It therefore leaves the combination $I^2 = \tilde \alpha^2 - 4 \tilde \beta$ invariant. This combination determines the special Galileon algebra, with 
 \begin{align}
  [ S_{\mu \nu} , S_{\rho \sigma} ] = \tfrac12 I^2 (\eta_{\rho (\mu} M_{\nu) \sigma} + \eta_{\sigma (\mu} M_{\nu) \rho})   \,.
 \end{align}
In contrast to the conformal case discussed before, not all special Galileon algebras are therefore related by contact transformations; instead, there are three inequivalent (orbits of) special Galileon algebras, depending on whether $I^2$ is positive (as happens for the Goldstone of linear diffeomorphisms discussed in the previous sections), zero or negative. Galilean duality acts as $SO(1,1)$ on this light-cone structure and leaves the norm invariant (whose overall magnitude can be absorbed in the scale $\Lambda$). 

A representative of the negative norm case is $\tilde \alpha = 0$ and $\tilde \beta$ positive, for which the covariant metric was found previously in \cite{Novotny}. The algebra of this orbit is of special unitary form instead. A geometric interpretation for this special unitary structure involves a 3-brane in a four-dimensional complex geometry with Kahler structure. This naturally gives rise to a metric of the form 
 \begin{align}
  g_{\mu \nu} = \eta_{\mu\nu} + \partial_{\mu} A^\rho \partial_\nu A_\rho \,.
 \end{align}
The subsequent inverse Higgs condition $A_\mu = \partial_\mu \phi$ then imposes the vanishing of the Kahler form on the 3-brane, restricting it to lie on a Lagrangian submanifold \cite{Novotny}. 

The amplitude for 2-to-2 scattering, being a physical observable, only depends on the invariant combination $I^2$. Similar to the DBI case, requiring a UV completion again leads to a positivity bound on this combination (after the addition of a mass term) \cite{Melville}. This picks out the $I^2$ positive orbit as being compatible with a standard UV completion. The fate of the $I^2 < 0$ case with special unitary transformations is therefore similar to that of anti-DBI, and can be related to the two times-aspect of its geometric embedding and $SU(3,1)$ structure.

The last possibility is the null case, with $I^2=0$, in which case the special Galileon transformations commute. Up to a contact transformation, this symmetry is equivalent to the quadratic shift symmetry of a scalar field with no accompanying diffeomorphism $(\tilde \alpha = \tilde \beta =0)$ that acts on the Minkowski metric. This transformation can be extended beyond the quadratic one, and for instance one finds (for $\tilde \alpha = 2 \tilde \beta =2$)
 \begin{align}
  \delta \phi &= c_{\mu\nu\rho} (x^\mu x^\nu x^\rho - 3 x^\mu \partial^\nu \phi \partial^\rho \phi - 2 \partial^\mu \phi \partial^\nu \phi \partial^\rho \phi) \,, \notag \\
  \chi_\mu & = 3 c_{\mu \nu \rho} (x^\nu + \partial^\nu \phi)(x^\rho + \partial^\rho \phi) \,, 
 \end{align}
for the cubic transformations with symmetric parameter $c_{\mu\nu\rho}$. All these are possible as the metric in the null case is actually flat, and does not contain a propagating scalar degree of freedom (the dynamical condition \eqref{unimod-gen} becomes trivial for this case). 

Interestingly, this case appears in the decoupling limit of massive gravity. This theory propagates three longitudinal modes, in addition to the two transversal ones. One can make these additional degrees of freedom manifest by introducing them as St{\"u}ckelberg fields in such a way as to make the full metric invariant under diffeomorphisms (which would otherwise be broken by the mass deformation)\cite{Schwartz} (see also \cite{Hinterbichler, deRham} for reviews). In the $\Lambda_3$ decoupling limit, keeping $\Lambda_3 = (m^2 M_{\rm Pl})^{1/3}$ fixed while $m \rightarrow 0$ and $M_{\rm Pl} \rightarrow \infty$, this yields
 \begin{align}
  g_{\mu \nu} = \eta_{\mu \nu} + \frac{2 \partial_\mu \partial_\nu \phi}{\Lambda_3{}^3} + \frac{ \partial_\mu \partial^\rho \phi \partial_\nu \partial_\rho \phi}{\Lambda_3{}^6} \,. \label{MG-metric}
 \end{align}
As follows from the discussion above, this metric transforms covariantly under commuting special Galileon transformations as well as higher-order generalisations. However, mass terms of the form $m^2 (\eta^{\mu \nu} g_{\nu \rho})^2$ always involve a reference metric (in this case the Minkowski one), and hence break these symmetries and propagate a scalar degree of freedom.

\section{Conclusions}

Recent studies have shown remarkable relations between different field theories, starting from the double copy relation between YM and GR and branching out to the colour-kinematics duality relations indicated in figure 1. These include many well-known theories such as NLSM, YM and GR, plus a more mysterious yet highly symmetric one: the special Galileon. 

We have shown that the special Galileon is a close relative of GR: it is the Goldstone associated to linear diffeomorphisms, similar to the relation between YM and NLSM. However, the Goldstone of the NLSM \eqref{gauge-configuration} corresponds to the longitudinal mode of massive YM, and mass terms $m^2 (A_\mu)^2$ give rise to the non-linear sigma-model. Instead, the Goldstone of the affine group of general relativity with metric \eqref{metric-aff} does not coincide with the longitudinal scalar mode of massive gravity\footnote{Upon completion of this manuscript, this point was also discussed in the context of amplitudes and their double copy \cite{Tolley} and was attributed to specific contact diagrams. It would be interesting to relate this to the difference between the metrics \eqref{metric-aff} and \eqref{MG-metric}.}. The latter metric \eqref{MG-metric} has vanishing curvature and realises commuting symmetries; however, there are no mass terms that respect these symmetries. The Goldstone metric \eqref{metric-aff} has non-vanishing curvature and non-commuting linear transformations; the SG dynamics is provided by the unimodularity condition, that is compatible with the special linear part of these transformations.

Both the special Galileon and general relativity are, in a very concrete sense, the end of the line: there are no interacting field theories for massless higher spins, and there are no scalar theories with softer limits \cite{softlimits1, softlimits2}, or equivalently higher-order symmetries \cite{Werkman}. Similarly, they are one of a kind: one cannot add interactions between massless spin-2 fields \cite{Henneaux}, and there are no multi-special Galileons known \cite{softlimits2, Elvang, Werkman}, in contrast to e.g.~the multi-DBI theories.

We have provided a novel formulation of the special Galileon in terms of the covariant metric \eqref{metric-aff} with non-linearly realised affine coordinate transformations\footnote{It would be interesting to investigate their curved counterparts; are there Anti-de Sitter analogues to the covariant metric and unimodularity condition, and do these allow for an intuitive formulation of $\phi$ as Goldstone of $SL(3,2)$ transformations \cite{AdS}?}. This suggests a number of similarities between the different theories on the bottom of figure 1. All are geometric and can be formulated in terms of metric that is given by $\eta + h$, $\eta + F$ and $\eta + \partial \partial \phi$ from left to right; note the increase in derivatives\footnote{This structure is mirrored by the scalar field theories at the right of figure 1; the covariant building blocks from BS to SG are $\phi^{a \bar a}$, $U^{-1} d U$ and $g_{\mu \nu}$, increasing in derivatives at every step.} as the spin decreases from GR to the SG. Relatedly, the field equations for these theories follow from the variation of the Einstein-Hilbert term for GR, the variation of the measure for Born-Infeld (BI) and the unimodularity condition for the SG. 

The relations between the GR and SG symmetries bring into sharper focus the absence of such an organising principle for BI. It is known from the amplitude perspective \cite{Elvang} that $U(1)$ gauge vectors do not allow for softer structures in the single-soft limit; relatedly, there are no non-trivial extensions of Poincare and the gauge symmetry \cite{Klein}. On the other hand, the Born-Infeld vector describes the worldvolume degrees of D-branes, and it has been suggested that they arise from the embedding in double geometry \cite{Watamura}. It will be interesting to see how the results described here can contribute to this issue. \\

\section*{Acknowledgments}

We are grateful to James Bonifacio, Kurt Hinterbichler, Austin Joyce, Ivan Kol{\'a}r, Shubham Maheshwari, Scott Melville, David Stefanyszyn and Andrew Tolley for stimulating discussions.

\end{document}